\documentclass[conference]{ieeeconf}
\let\labelindent\relax
\IEEEoverridecommandlockouts
\usepackage{cite}
\usepackage{amssymb,amsfonts,amsmath,graphicx}
\usepackage{multirow}
\usepackage{amssymb}
\usepackage{xcolor}
\usepackage{textcomp}
\usepackage{float}
\usepackage{placeins}
\usepackage[misc]{ifsym}
\def\BibTeX{{\rm B\kern-.05em{\sc i\kern-.025em b}\kern-.08em
    T\kern-.1667em\lower.7ex\hbox{E}\kern-.125emX}}
\usepackage{enumitem}
\setlist{nosep, leftmargin=14pt}

\usepackage{mwe} 

\newcommand{\tabincell}[2]{\begin{tabular}{@{}#1@{}}#2\end{tabular}}
\begin{document}

\title{MMMNA-Net for Overall Survival Time Prediction of Brain Tumor Patients\\
\thanks{*Wen Tang and Haoyue Zhang contributed equally. }
\thanks{Wen Tang, Haoyue Zhang, Pengxin Yu, Han Kang, Rongguo Zhang are with InferVision Medical Technology Co.,Ltd., Beijing, China. {twen,shaoyue,ypengxin,khan,zrongguo}@infervision.com}
\thanks{Haoyue Zhang is with Department of Bioengineering, University of California, Los Angeles, Los Angeles, CA, USA. zhanghaoyue@g.ucla.edu}
\thanks{©20XX IEEE.  Personal use of this material is permitted. Permission from IEEE must be obtained for all other uses, in any current or future media, including reprinting/republishing this material for advertising or promotional purposes, creating new collective works, for resale or redistribution to servers or lists, or reuse of any copyrighted component of this work in other works. }
}
\author{Wen~Tang*,~
        Haoyue~Zhang*,~
        Pengxin~Yu,~
        Han~Kang,~
        and~Rongguo~Zhang}%

\maketitle              

\begin{abstract}
Overall survival (OS) time is one of the most important evaluation indices for gliomas situations. Multi-modal Magnetic Resonance Imaging (MRI) scans play an important role in the study of glioma prognosis OS time. Several deep learning-based methods are proposed for the OS time prediction on multi-modal MRI problems. However, these methods usually fuse multi-modal information at the beginning or at the end of the deep learning networks and lack the fusion of features from different scales. In addition, the fusion at the end of networks always adapts global with global (eg. fully connected after concatenation of global average pooling output) or local with local (eg. bilinear pooling), which loses the information of local with global. In this paper, we propose a novel method for multi-modal OS time prediction of brain tumor patients, which contains an improved non-local features fusion module introduced on different scales. Our method obtains a relative 8.76\% improvement over the current state-of-art method (0.6989 vs. 0.6426 on accuracy). An extra testing demonstrates that our method could adapt to the situations with missing modalities. The code is available at https://github.com/TangWen920812/mmmna-net.
\end{abstract}
\section{Introduction}
Brain tumor is the most common and difficult-to-treat malignant neurological tumor with the highest mortality rate. Glioma accounts for about 70\% of primary brain tumors in adults~\cite{ref_article1}. Overall survival (OS) time is a main consideration when assessing patient situation and making treatment plans. Thus, timely and accurate OS time prediction of patients with glioma is of great clinical importance.
Multi-modal Magnetic Resonance Imaging (MRI) scans, such as native (T1), T1 contrast-enhanced (T1Ce), T2-weighted (T2), and fluid attenuated inversion recovery (FLAIR) scans are commonly used in the study of the prognosis OS time for patients with glioma. Recently, there are mainly two types of methods based on deep learning algorithms proposed for survival prediction using multi-modal MRI scans. In the first type of methods, different modalities are concatenated in the dimension of channel as input. For example, Nieet al.~\cite{ref_article2} use a 3D deep learning model to extract multi-modal multi-channel features and then feed them into a support vector machine classifier for OS prediction. Such kind of approaches combine multi-modal images at the beginning and then extract features from multi-modal input. Yet, these methods mainly rely on the effectiveness and robustness of the network feature extraction on complicated input, and does not perform apriori merging of deep-level features. It may lead to poor performance when only a limited number of training samples are available. Another type of methods combine multi-modal features at the end of feature processing. Feng et al.~\cite{ref_article3} propose a linear model for survival prediction by concatenating multi-modal imaging features extracted by backbone networks. Tao Zhou et al.~\cite{ref_article4} use non-shared networks to extract multi-modal features and then use bilinear pooling to combine features. Such methods concatenate multi-modal features at the end of the network. Thus, low-level multi-modal features are not fused in early stage. In addition, feature fusion performed in these methods is only based on the average ~\cite{ref_article3} or the same position of the last layer features ~\cite{ref_article4} and does not reflect the local and global relevance. 

To better leverage the features extracted from multi-modal images, we propose an end-to-end multi-modal multi-channel multi-scale non-local attention network (MMMNA-Net) for OS time prediction. Multi-modal shared backbone is used to automatically extract low-level and high-level features from different modality inputs. Then we apply improved non-local attention modules to fuse features at each level. At the highest level of features before average pooling, a fully connection layer is used instead of global average pooling to enhance the relevance of each modality. Compared with conventional classification-based OS prediction methods, MMMNA-Net can fuse both high-level and low-level features extracted from different modalities with relevance between local and global. Results show that the MMMNA-Net outperforms several other classification-based OS prediction methods. In addition, even though our model is trained on four modalities, it can also be applied to data without all four modalities.

\begin{figure}
\begin{center}
\includegraphics[width=\columnwidth]{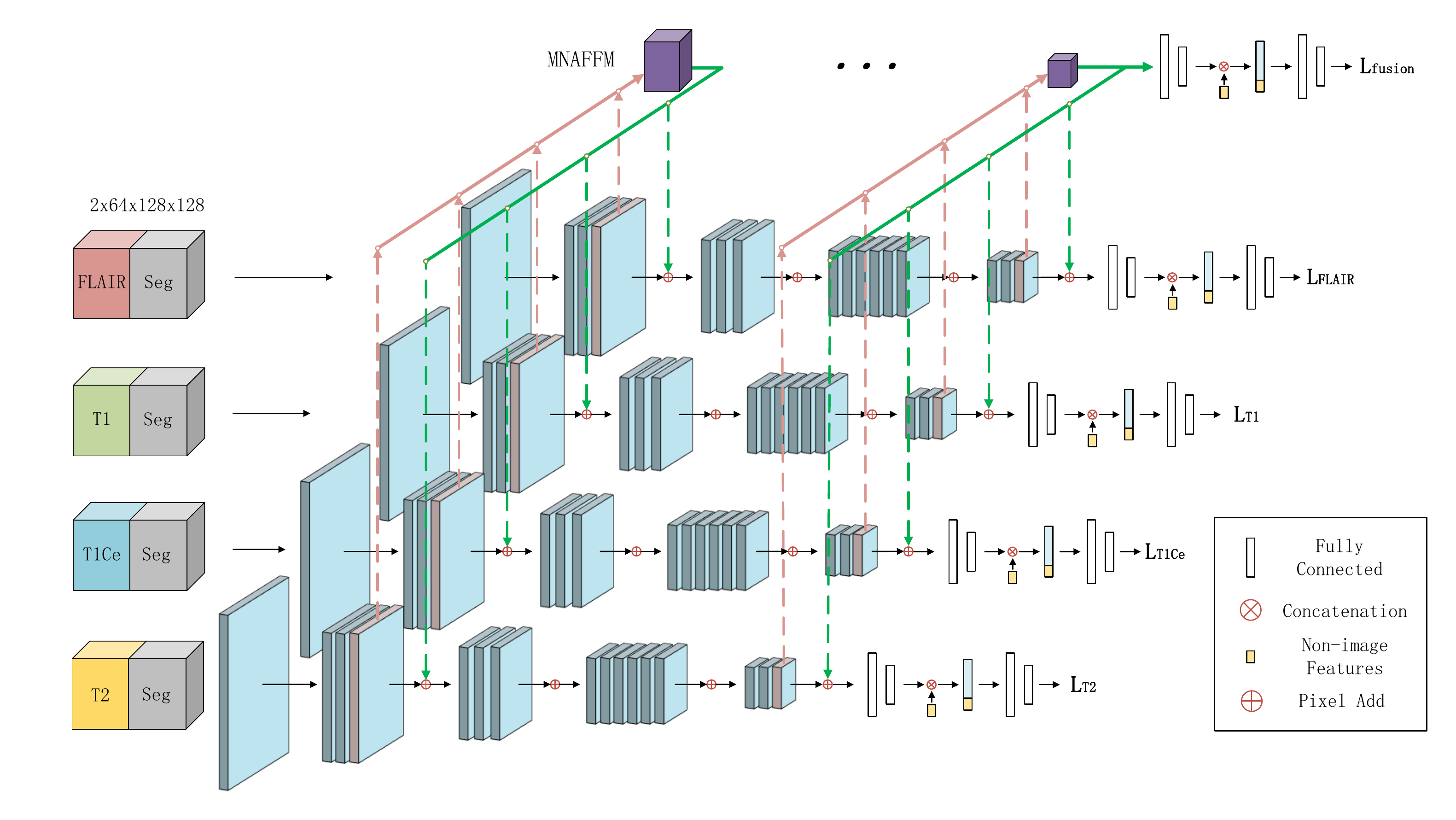}
\end{center}
\caption{Overall structure of Multi-modal Multi-channel Multi-scale Non-local Attention Network.} \label{fig1}
\end{figure}

\section{Method}
Our MMMNA-Net consists of two parts: the multi-modal multi-channel shared network backbone and the multi-scale non-local attention feature fusion module. We detail these two parts and show how to combine them together as the proposed MMMNA-Net below.

\subsection{Multi-modal multi-channel shared network}
Convolutional Neural Networks (CNNs) have been widely applied to perform visual tasks. 3D CNNs are commonly used on medical image processing as spacial information of voxels could be better used in 3D CNNs compared to 2D CNNs. Considering that training 3D CNNs requires more data, we build our shared backbone based on 3D ResNet18 \cite{ref_article5} structure with modifications to reduce the amount of parameters to avoid overfitting. The stride size of the first $7\times7\times7$ kernel convolution is changed from $2\times2\times2$ to $1\times2\times2$ to avoid the possible domain shift caused by the difference between inter-slice and intra-slice resolution. As shown in Figure \ref{fig1}, we concatenate the segmentation annotations with each modality as input to make sure the 3D CNNs would pay more attention on the lesion area while not ignoring other areas which may be useful for OS time prediction. Each modality volume concatenated with segmentation annotations would be sent into the shared parameter backbone to extract features after data preprocessing.

\subsection{Multi-scale non-local attention feature fusion module (MNAFFM)}

\begin{figure}
\includegraphics[width=\columnwidth]{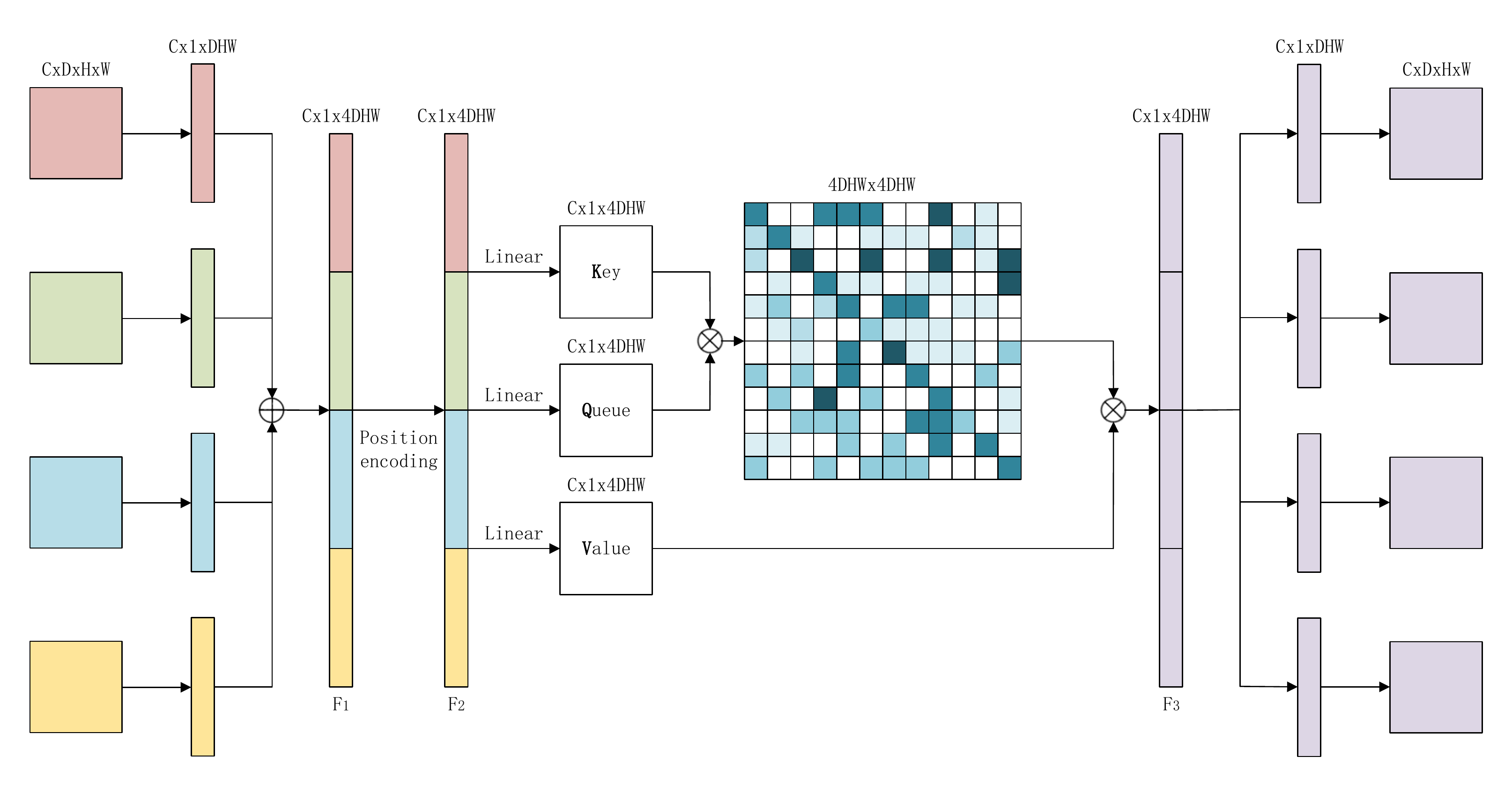}
\caption{Non-local multi-modal fusion Flowchart} \label{fig2}
\end{figure}

Non-local~\cite{ref_article6} is an attention-based method first proposed in the field of natural language processing. It is applied to deep learning algorithms for image processing in order to increase receptive field and the relevance between local and global information. 
As shown in Figure \ref{fig2}, we introduce the non-local method to the fusion of multi-modal features. We firstly spread the features of all modalities and concatenate them as feature $F_{1}$. Then, we add position information to $F_{1}$ as shown in BERT~\cite{ref_article7} and get feature $F_{2}$. These position features are encoded by equation \ref{eq1} and equation \ref{eq2}, in which $d_{model}$ is the dimension of modal features. Such method is proved useful according to the experiments by Dosovitskiy et al. \cite{ref_article8}. Then the non-local is performed on $F_{2}$ as a self-attention module. Therefore, each voxel in feature map is related to all voxels including itself. Finally, fused features $F_{3}$ are split equally into 4 parts and reshaped back to the original input shape. However, due to the memory limitation, non-local is usually used at the end of network. Given this situation, we introduce an improved non-local approach, the self-attention module in Linformer \cite{ref_article9}, into our multi-modal network to reduce the complexity from $O(n^2)$ to $O(n)$. The lowered computational cost allows us to use non-local attention fusion module on different scales of features.

\begin{equation}\label{eq1}
PE_{(pos/2 == i)} = sin(\frac{pos}{10000^{2i/d_{model}}}) 
\end{equation} 
\begin{equation}\label{eq2}
PE_{((pos-1)/2 == i)} = cos(\frac{pos}{10000^{2i/d_{model}}})
\end{equation} 

\begin{figure}
\includegraphics[width=\columnwidth]{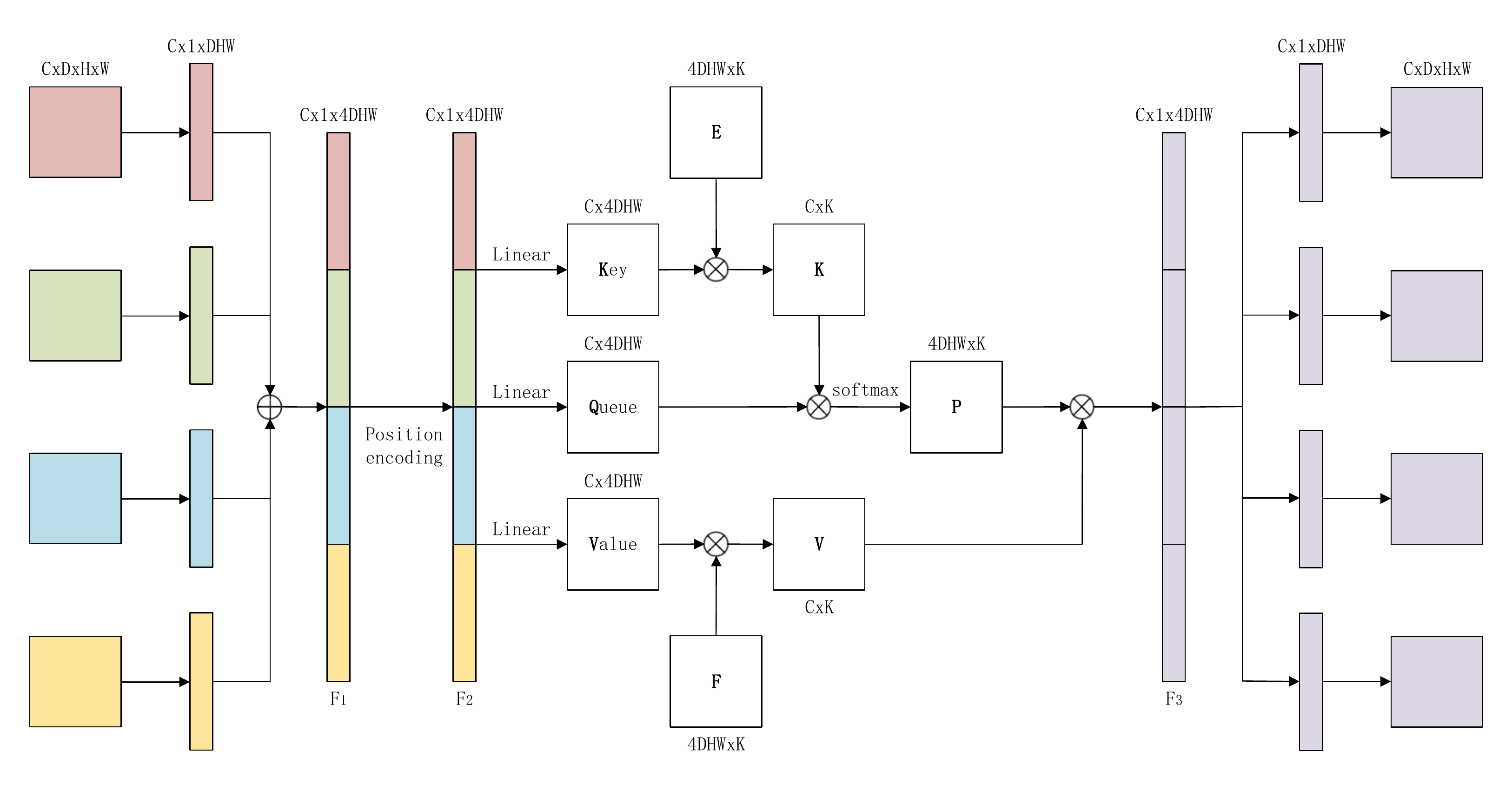}
\caption{Multi-scale features fusion module} \label{fig3}
\end{figure}

Figure \ref{fig3} shows the improved non-local attention feature fusion module we used in our MMMNA-Net. The features are first processed same as the non-local attention module. Unlike in the non-local attention module, after we obtain $Key$,$\;Queue$ and $Value$ from concatenated features, no attention maps are calculated. Instead, we use $E$ and $F$, two learnable matrix with random initialization, to reduce the dimension of $Key$ and $Value$. As shown in equation \ref{eq3}, we first conduct a linear transformation on $Key$ and $Value$ by multiplying them with $E$ and $F$ to get $K$ and $V$ with reduced dimension, respectively. Then, we multiply $Queue$ with $K$ and then perform softmax on the result to get $P$, which is the approximate expression of global attention map. Finally, we multiply $P$ with $V$ and get fused features. Similar kind of dimension reduction is proved to be effective in \cite{ref_article9}. Fused features $F_{3}$ are split equally into 4 parts and reshaped back to the original input shape. 
\begin{equation} \label{eq3}
Fused \; features = softmax(Queue \times Key \times E) \times (Value \times F)
\end{equation}

\subsection{Multi-modal multi-channel multi-scale non-local attention network}
We apply MNAFFM on different scales of multi-modal features. We add MNAFFM on positions before each max-pooling layer except for the first one due to the size of which is too big and shallow to fuse. The size of the input of each module is $32\times32\times32\times4$, $16\times16\times16\times4$, $8\times8\times8\times4$ and $4\times4\times4\times4$, respectively. Figure \ref{fig1} shows the overall structure of our MMMNA-Net. The outputs of MNAFFM are added to the original input of the module and form a residual structure which would accelerate the training process and the features of greater importance would be identified during this process. After the process of the last MNAFFM, we use a branch-specific fully connected layer instead of global average pooling to perform a weighted average pooling for a stronger relevance of pixels in each branch (fusion, FLAIR, T1, T1Ce and T2). 
Notably in the fusion branch, feature $F_{1}$ is added with fusion feature $F_{3}$ as the input for the weighted average pooling layer.
After that, non-image information is added at the end of the whole network before the last branch-specific fully connected classification layer. Inspired by the MMNet \cite{ref_article4}, loss of each modality (T1, T1Ce, T2, FLAIR) is calculated to assist the convergence. Focal loss is used as the classification loss due to the imbalanced sample issue in training data. Loss of MMMNA-Net can be obtained as:

\begin{equation}
Focal \; loss = -\alpha * (y) * (1 - p) ^ \gamma * log(p) - (1-\alpha) * p^\gamma*log(1-p)
\end{equation}
\begin{equation}
Total \; Loss = (L_{T1} + L_{T1Ce} + L_{T2} + L_{FLAIR}) * \lambda + L_{fusion}
\end{equation}
Where $\alpha$ is the weighting factor to mitigate class imbalance and $\gamma$ is the focusing parameter to adjust for easy samples. The total loss is accumulated by weighted summation of losses for each modality and a fusion loss. 

\section{Experiments}
\subsection{Data description}
The BraTS2020 \cite{ref_article10}\cite{ref_article11} is collected for the segmentation of intrinsically heterogeneous brain tumor sub-regions and contains multi-institutional pre-operative MR scans. In this dataset, each patient includes 1) T1, 2) T1Ce, 3) T2, 4) FLAIR volumes, and 5) lesion segmentation annotation. We use 237 subjects with survival information (in days) to perform OS time prediction.

According to the evaluation metric \cite{ref_article12} from the BraTS 2020, the continuous survival time was divided into three categories: (1) short-term survivors (i.e., $\leq$10 months) (2) mid-term survivors (i.e., between 10 and 15 months), and (3) long-term survivors (i.e., $\geq$15 months). For our OS prediction model, we resample the original volume data from $155\times253\times253$ to $64\times128\times128$ using bicubic interpolation. We also resample the segmentation annotation to $64\times128\times128$ by nearest neighbor interpolation for the concatenation of volume data and annotation. The tumor size is non-image information related to brain status, and could also influence the prediction performance. Thus, we also adapt similar approach as MMNet to process it: for every type of tumor, the size of each tumor is calculated using: $s_{i} =n_{i}/ \textstyle\sum_{i}n_{i}(i = 1, 2, 4) $, where $n_{i}$ is the voxel number of the i-th type of tumor. The total size is also calculated by $s_{total} =\frac{\textstyle\sum_{i}n_{i}}{ n_{non-zero}} $, where $n_{non-zero}$ is the total voxel number of the non-zero elements in the original MR volume. Further, the age information (denoted as “$s_{age}$”) is added using $s_{age} =\frac{age}{100} $ into the additional feature vector. 

\subsection{Overall implementations}

A 10-fold cross-validation strategy is adopted in experiments for performance evaluation with four metrics, including accuracy, precision, recall, and F-score \cite{ref_article13}. The precision and recall are calculated with the one-class-versus-all-other-class strategy, and F-score is calculated by equation \ref{eq6}.
\begin{equation} \label{eq6}
F\mbox-score = \frac{2 * precision * recall}{precision + recall}
\end{equation}

We compare our MMMNA-Net with the following methods: 1) 3D CNN concatenation method. We concatenate all modalities volume and corresponding segmentation annotation as input, and put them into a 3D CNN (3D ResNet18). 2) 3D CNN fusion method. We use specific 3D CNN (3D ResNet18) for each modality, and concatenate features after global average pooling. 3) 2D CNN concatenation method. We project the 3D volume same as MMNet, and concatenate them into a 2D CNN (ResNet34) as one input. 4) 2D CNN fusion method. We use specific 2D CNN (ResNet34) for each modality, and concatenate the features after global average pooling. 5) MMNet. We replicate this method based on the MICCAI 2020 \cite{ref_article4}. 6) 3D shared CNN with non-local fusion. It is similar to our method but with only the last layer non-local feature fusion module. 7) 3D modality specified CNN with non-local fusion. Comparing to 6), we use modality specified 3D CNN instead of a shared one. 8) 3D modality specified MMMNA-Net. Our proposed method with modality specified 3D CNN backbone. Noted that this method takes too much memory to achieve under current experiment setting.

All these methods listed above are implemented using PyTorch, and all models are trained using Adam optimizer with $10^{-4}$ learning rate. The weight decay rate is set to $10^{-4}$ and the maximum epoch is 200 with early stopping. Models are trained on 4 NVIDIA 1080 GPUs. The batch sizes are different across models based on memory cost. Besides, training data are augmented by vertical and horizontal flipping, and rotation. The trade-off parameter is set to $\lambda = 0.25$.

\subsection{Model comparison}
Table \ref{tab1} shows the prediction performance of all methods. The accuracy of the models are compared using the McNemar test. $p<0.05$ is considered statistically significant. All the differences are statistically significant except for 3D shared CNN with non-local fusion. Our proposed method outperforms all the comparing methods regarding the four metrics. It indicates the effectiveness of our method. Both variants of our proposed method, 3D modality specified CNN with non-local fusion and 3D and 3D shared CNN with non-local fusion, also exceed the current state-of-art method MMNet, showing that non-local fusion module is beneficial for performance improvement. MMMNA-Net improves the performance slightly compared to the 3D shared CNN with non-local fusion. Based on the training process, it could be due to the fact that MMMNA-Net has much more parameters than the 3D shared CNN with non-local fusion, and therefore is hard to train on a small dataset. 

\FloatBarrier
\begin{table}[h]
\caption{Model comparison result table}\label{tab1}
\begin{center}
\resizebox{\columnwidth}{!}{
\begin{tabular}{c|c|c|c|c}
\hline
Method & Accuracy & Recall & Precision & F-score \\
\hline
\tabincell{c}{3D CNN \\concatenation}& $0.6049\pm0.0703$ & $0.5243\pm0.0536$ & $0.5107\pm0.0712$ & $0.5166\pm0.0603$ \\
\hline
3D CNN fusion& $0.5765\pm0.0552$ & $0.5068\pm0.0624$ & $0.4902\pm0.0556$ & $0.4966\pm0.0519$\\
\hline
\tabincell{c}{2D CNN \\concatenation}& $0.6143\pm0.0611$ & $0.5378\pm0.0538$ & $0.5202\pm0.0664$ & $0.5246\pm0.0431$ \\
\hline
\tabincell{c}{2D CNN fusion}& $0.5938\pm0.0622$ & $0.5168\pm0.0584$ & $0.5102\pm0.0484$ & $0.5121\pm0.0474$ \\
\hline
MMNet& $0.6426\pm0.0334$ & $0.5478\pm0.0278$ & $0.5502\pm0.0409$ & $0.5484\pm0.0294$ \\
\hline
\tabincell{c}{3D modality \\ specified CNN \\with non-local fusion}& $0.6634\pm0.0431$ & $0.6430\pm0.0306$ & $0.6384\pm0.0585$ & $0.6401\pm0.0432$\\
\hline
\tabincell{c}{3D shared CNN \\with non-local fusion}& $0.6977\pm0.0369$ & $\mathbf{0.6784\pm0.0490}$ & $0.6704\pm0.0748$ & $0.6592\pm0.0679$ \\
\hline
\tabincell{c}{3D modality specified \\MMMNA-Net}& - & - & - & - \\
\hline
MMMNA-Net& $\mathbf{0.6989\pm0.0371}$ & $0.6778\pm0.0496$ & $\mathbf{0.6802\pm0.0557}$ & $\mathbf{0.6613\pm0.0539}$ \\
\hline
\end{tabular}
}
\end{center}
\end{table}
\FloatBarrier

\subsection{Missing modality experiments}
Considering the actual clinical situation that not all patients have four modalities of image, we also validate our model on testing data contain three or less modalities. We substitute the missing modalities with FLAIR modality, which is commonly used in OS time prediction. As shown in Table \ref{tab2}, our model performs well when different modalities are missing. On testing data containing two or less modalities, our model still outperforms all other comparison methods. All the differences are statistically significant. It may be because the MMMNA-Net learns multi-modal feature expression with fewer modalities during the training. When comparing three modalities settings, we notice that when T1 and T1Ce are used with FLAIR, the performance is lower than either T1 with T2 or T1Ce with T2, suggesting that T2 contrast with T1 weighted sequences may provide useful information for the model to learn. One interesting finding is that FLAIR+T1+T2 shows slightly better performance than FLAIR+T1Ce+T2. Although T1Ce shows the enhancing tumor visually, it does not contribute more to the prediction of OS comparing to T1. More experiments with larger dataset is necessary to re-examine this observation.

\FloatBarrier
\begin{table}[h]
\caption{Model performance with missing modality}\label{tab2}
\begin{center}
\resizebox{\columnwidth}{!}{
\begin{tabular}{cccc|c|c|c|c}
\hline
FLAIR, & T1, & T1Ce, & T2 & Acc & Recall & Precision & F-score \\
\hline
$\checkmark$ & & & & $0.6185\pm0.0631$ & $0.5707\pm0.0830$ & $0.5519\pm0.1069$ & $0.5389\pm0.0940$ \\

$\checkmark$ & $\checkmark$ & & & $0.6007\pm0.0714$ & $0.5569\pm0.0893$ & $0.5493\pm0.1106$ & $0.5287\pm0.1000$ \\

$\checkmark$ & & $\checkmark$ & & $0.6228\pm0.0222$ & $0.5709\pm0.0691$ & $0.5532\pm0.1178$ & $0.5358\pm0.0800$ \\

$\checkmark$ & & & $\checkmark$ & $0.6562\pm0.0715$ & $0.6216\pm0.1012$ & $0.6402\pm0.1300$ & $0.6048\pm0.1079$ \\

$\checkmark$ & $\checkmark$ & $\checkmark$ & & $0.6353\pm0.0302$ & $0.5905\pm0.0667$ & $0.6005\pm0.1129$ & $0.5602\pm0.0794$ \\

$\checkmark$ & $\checkmark$ & & $\checkmark$ & $0.6687\pm0.0735$ & $0.6412\pm0.1020$ & $0.6637\pm0.1305$ & $0.6212\pm0.1120$ \\

$\checkmark$ & & $\checkmark$ & $\checkmark$ & $0.6611\pm0.0562$ & $0.6217\pm0.0884$ & $0.6337\pm0.1216$ & $0.6034\pm0.0936$ \\

$\checkmark$ & $\checkmark$ & $\checkmark$ & $\checkmark$ & $\mathbf{0.6989\pm0.0371}$ & $\mathbf{0.6778\pm0.0496}$ & $\mathbf{0.6802\pm0.0557}$ & $\mathbf{0.6613\pm0.0539}$ \\
\hline
\end{tabular}
}
\end{center}
\end{table}
\FloatBarrier

\section{Conclusion}
We propose a multi-modal multi-channel multi-scale non-local attention network to solve the OS time prediction problem on multi-modal MR scans from BraTS2020 dataset. An improved non-local attention module is used to assist the fusion of different scales features in the network. We compare our method with several multi-modal methods including a novel method from MICCAI 2020. Our single-scale (3D shared CNN with non-local fusion) and multi-scale networks (MMMNA-Net) outperform all other methods. According to the results, our multi-scale network outperforms slightly to the single-scale network. A bigger dataset could be used to conduct more experiments on these two models in future work. In addition, our model is trained with four modalities and perform well on data containing three or less modalities. It shows that our proposed method could function well even with missing modalities.

\section{Acknowledgement}
No funding was received for this study. The
authors have no relevant financial interests to disclose.

%
%
%
\bibliographystyle{splncs04.bst}
\bibliography{ref.bib}
\end{document}